# Systematic study of disorder induced by neutron irradiation in MgB$_2$ thin films

V. Ferrando, I. Pallecchi, C. Tarantini, D. Marré, M. Putti and C. Ferdeghini

*LAMIA-CNR-INFM and Università di Genova, via Dodecaneso 33, 16146 Genova Italy*

F. Gatti

*Università di Genova, via Dodecaneso 33, 16146 Genova Italy*

H. U. Aebersold and E. Lehmann

*Paul Scherrer Institut, CH-5232 Villigen, Switzerland*

E. Haanappel,

*Laboratoire National des Champs Magnétiques Pulsés, CNRS-UPS-INSA, Toulouse, France*

I. Sheikin,

*GHMFL, MPI-FKF/CNRS, 28 Avenue des Martyrs, BP 166, 38042 Grenoble Cedex 9, France*

P. Orgiani* and X. X. Xi

*The Pennsylvania State University, University Park, PA 16802, USA*

**Abstract**

The effects of neutron irradiation on normal state and superconducting properties of epitaxial magnesium diboride thin films are studied up to fluences of $10^{20}$ cm$^{-2}$. All the properties of the films change systematically upon irradiation. Critical temperature is suppressed and, at the highest fluence, no superconducting transition is observed down to 1.8 K. Residual resistivity progressively increases from 1 to 190 µΩcm; *c* axis expands and then saturates at the highest damage level. We discuss the mechanism of damage through the comparison with other damage procedures. The normal state magnetoresistivity of selected samples measured up to high fields (28 and 45T) allows to determine unambiguously the scattering rates in each band; the crossover between the clean and dirty limit in each sample can be monitored. This set of samples, with controlled amount of disorder, is suitable to study the puzzling problem of critical field in magnesium diboride thin films.

The measured critical field values are extremely high (of the order of 50T in the parallel direction at low fluences) and turns out to be rather independent on the experimental resistivity, at least at low fluences. A simple model to explain this phenomenology is presented.

*Presently at CNR-INFM Supermat, Via Allende, I-84081 Baronissi (Sa), Italy

# I. Introduction

After five years since the discovery of superconductivity in magnesium diboride [1], the effect of disorder in this material is one of the points still under discussion in the scientific community, either to understand the fundamental physics of this system or to improve its performance in magnetic field. It is well known that the Fermi surface of $MgB_2$ consists of two sets of bands: nearly two dimensional anisotropic $\sigma$ bands, mostly located in the boron planes and strongly coupled with $E_{2g}$ phonon mode, and three dimensional isotropic $\pi$ bands, which are distributed over all crystalline cells[2]. Due to the very weak interband scattering, $\sigma$ and $\pi$ bands can be described as two channels conducting in parallel. Therefore defects introduced in magnesium planes should not affect $\sigma$ channel, as defects in boron planes mainly influence $\sigma$ channel. In this framework, disorder introduced by chemical substitutions has been intensively studied especially in polycrystalline bulk samples and single crystals; in particular, only aluminium [3-5] and carbon[6,7] have shown a clear evidence of entering the $MgB_2$ structure. Both of them dope the system with electrons, filling up the $\sigma$ bands causing a suppression of critical temperature. Irradiation by ions, protons or neutrons is another possible route to systematically change the disorder in this material without change of charge in the system; it produces large scale and point defects, which may improve superconducting properties of magnesium diboride. Different studies on irradiation have been already reported in literature. In polycrystalline bulk and in single crystals, protons produce an increase of irreversibility field and an improved magnetic field dependence of $J_c$, with a small decrease of $T_c$[8,9]. Neutron irradiation, performed both on single crystals and on different kind of polycrystalline samples, showed a similar enhancement of the superconducting properties, but leading to a strong suppression of $T_c$ at high fluences, fully recovered by annealing[10-14]. Also the merging of the gaps was recently observed on neutron damaged bulk samples[15]. In thin films, the effect of disorder is even more complicated and still not well understood. Nominally clean films grown by Pulsed Laser Deposition and other techniques[16,17], usually do not have an optimal and well reproducible critical temperature. Furthermore, the problem of critical fields in $MgB_2$ thin films is puzzling and still unsolved. In literature, extremely high values were observed in nominally clean films prepared by different techniques[16-18], even when residual resistivity is extremely low. In addition, films with resistivity that differs by one order of magnitude can present very similar critical field values[19], suggesting the presence of some intrinsic disorder unintentionally introduced during the growth process. This makes the systematic study of disorder hard. Up to now, Hybrid Physical Vapor Deposition (HPCVD) is the only technique which produces epitaxial high quality films, with reproducible $T_c$ up to 41K and $\rho_0$ even lower than those of single crystals[18]; also these samples have $H_{c2}$ values higher than bulks, but they can anyway be considered as clean systems in which disorder can be introduced in a controlled way. Recently, Gandikota et al.[20-21] have studied the effect of damage by α particles on these films; they showed that α-particle irradiation progressively increases normal state resistivity and reduces $T_c$ down to 7 K without the saturation at 20 K theoretically expected in $MgB_2$ [22]. In this paper, we present our results on a series of films grown by HPCVD that have been neutron irradiated at different fluences up to $10^{20}$ cm$^{-2}$. The effects of this kind of damage on structural, normal state and superconducting properties will be discussed. In particular, we will try to shed light on the puzzling problem of the high $H_{c2}$ in thin films, thanks to the availability of a controlled series of disordered samples. We will introduce a simple model to explain the independency of critical fields on resistivity.

**II. Experimental**

The films were grown at The Pennsylvania State University by HPCVD on 5 x 5 mm$^2$ silicon carbide substrates, following the standard procedure described in [18]. All the samples were deposited nominally under the same conditions, at 720 °C and the estimated thickness was about 2000 Å, corresponding to an expected critical temperature of about 41.3K [23]. In order to avoid contamination by oxygen or humidity, which have dramatic effects on the superconducting properties of this kind of films, each film was sealed under high vacuum in a small quartz ampoule just after deposition; this setup allowed to perform irradiation also for long times without compromising the initial quality of the samples. Neutron irradiation was carried out at the spallation neutron source SINQ of Paul Sherrer Institut (PSI) in Villigen (Zurich), which has a thermal and fast neutrons flux density of $1.6 \cdot 10^{13}$ cm$^{-2}$s$^{-1}$ and $10^{10}$ cm$^{-2}$s$^{-1}$ respectively. As was recently shown in the case of bulk samples[10,12,14], neutrons produce defects by the neutron capture reaction of $^{10}$B atoms, which decay in a 0.84 MeV $^7$Li nucleus and in a $\alpha$ particle. The $\alpha$ particle energy depends on the final state of Litium: in particular it is 1.7 MeV or 1.47 MeV if the decay is on the fundamental state of Li (7% of the cases) or in the first excited state of Li nucleus respectively. This capture reaction has a huge cross section, so that in samples prepared with natural boron, where $^{10}$B is about 20% of the total, the penetration depth is few hundreds of $\mu$m, leading to an inhomogeneous distribution of defects in massive samples. In contrast, in thin films (2000 Å thick), the damage is expected to be uniform. Eleven different samples (IRR10-IRR50 in the following) have been irradiated at thermal neutron fluences from $6 \cdot 10^{15}$ to $10^{20}$ cm$^{-2}$ in three different runs. The irradiation levels have been chosen in order to be in the same range of the irradiated bulk samples of ref. 12; for the heaviest irradiated film, the required irradiation time was about one month.

The structural characterization of the films was performed by standard X-ray technique in Bragg-Brentano geometry; Rietveld analysis has been used to extract cell parameters.

To evaluate upper critical fields, electrical resistivity measurements were performed at the Grenoble High Magnetic Field Laboratory (GHMFL) by a standard four probe technique in static magnetic field up to 28 T, oriented both parallel and perpendicular to the *ab* planes. Some $H_{c2}$ curves were completed at low temperature at the 300 ms 60 T pulsed field facility at the Laboratoire National des Champs Magnétiques Pulsés (LNCMP) in Toulouse.

Normal state magnetoresistivity was measured both at GHMFL (up to 28T) and at LNCMP (up to 45T) in the less disordered samples: also in this case the magnetic field was applied perpendicular and parallel to the *ab* planes.

**III. Sample characterization**

Figure 1 shows (001) and (002) peaks for the clean and for selected damaged films. Upon irradiation, the intensity decreases progressively and the peaks slightly broaden, remaining anyway quite narrow even at the highest fluence ($10^{20}$ cm$^{-2}$). This is an indication that irradiation does not destroy the crystalline quality of the samples. From the FWHM of the (00*l*) one can calculate the size of the grains along the *c* axis. It passes from about 550 Å for IRR10, 20 and 30 to about 350 Å in the most damaged films, suggesting that irradiation has an influence on correlation length between the boron planes, as proposed in ref. 13, where anisotropy of the in-plane and out-of-plane correlation lengths has been found. The calculated *c* axis as a function of fluence are reported in upper panel of figure 2; similarly to what observed in bulk samples, the *c* axis expands by more than 1% from the unirradiated film up to IRR40 sample (thermal neutron fluence 9·10$^{18}$ cm$^{-2}$) and then it shows a slight decrease up to the heaviest irradiation level. This feature is still not well explained; also in bulk samples anyway, a saturation of the *c* axis was measured even though at higher fluences[14].

Table I summarizes the properties of the irradiated films series: thermal neutron fluence, critical temperature $T_c$ (defined as 50% of residual resistivity) and transition width $\Delta T_c$ (10-90% of normal state resistivity) estimated from electrical four probe measurements, residual resistivity ratio (RRR), measured residual resistivity $\rho_{exp}$, and $\Delta\rho=\rho(300K)-\rho(42K)$. Starting from the unirradiated sample IRR0 and increasing neutron fluence, the critical temperature decreases and the most irradiated sample, IRR50, does not show any superconducting transition down to 1.8K. The superconducting transitions slowly broaden remaining anyway quite sharp up to 7 10$^{17}$ cm$^{-2}$, then $\Delta T_c$ has a maximum in IRR 40 sample (fluence 9.5 10$^{18}$ cm$^{-2}$), decreasing again in the most irradiated superconducting film. In this sample, $\Delta T_c$ is extremely narrow, 0.2 K, even though the critical temperature is less than 3 K. A similar behaviour was observed in irradiated bulks[14]. The measured residual resistivity is about 1 µΩcm in the less irradiated samples but becomes about 2500 µΩcm in the most damaged one; RRR rapidly decreases down to about 1. According to Rowell [24], $\Delta\rho$ can be considered as an indication of the effective connectivity between grains and therefore of the reliability of the measured resistivity. As shown in Table I, $\Delta\rho$ in general is almost constant and varies between 8 and 11 in the majority of the films, indicating that the grains are well connected and experimental $\rho_{exp}$ is a realistic estimation of the intrinsic ρ. Only four samples (IRR35, 36, 45 and 50) show a larger $\Delta\rho$ along with a high resistivity. It is worth noting that these films have been irradiated in the same run and hence it is possible that some small differences in the times they remained in air occurred, although we followed the same procedure as for the previous runs. Therefore, in order to have reasonable and homogeneous resistivity values, the experimental resistivities of the whole series of samples has been normalized by using $\rho_{corr}=\rho_{exp}\Delta\rho_{theo}/\Delta\rho$ to the value $\Delta\rho_{theo}= 7.6$ µΩcm reported in ref. 24 for HPCVD clean samples. The corrected values are reported again in Table 1. In figure 2, the fluence dependence of $T_c$ and corrected resistivity along with the crystallographic parameters are reported. At low fluence, of the order of $10^{15}$-$10^{16}$ cm$^{-2}$, the properties of the films are only slightly influenced by irradiation; in the less irradiated film, the critical temperature is reduced by half a degree and the resistivity only slightly increases with respect to the unirradiated sample. Then the $T_c$ and ρ versus fluence curves become steeper and the dependence on irradiation stronger. It is worth noting that although the data of figure 2 are referred to samples irradiated in three different runs, they all follow the same trend. This is an indication that the effects of neutron irradiation are able to systematically modify the properties of the films.

Recently, Gandikota et al.[20] reported on the irradiation of thin films by 2 MeV α particles. Actually, the results of this experiment can be compared with ours, considering that the $^{10}$B neutron capture reaction products are heavy charged particles, i.e. Li and He nuclei with energies of 0.84 and 1.47

MeV (93% of the cases), that irradiate the film itself. The main difference between the two methods of irradiation is that in our case the α particle emission is coupled with a Li nucleus, whose linear energy transfer (LET) to the crystal is considerably higher than α particle. In order to evaluate the effectiveness in damaging the crystal, we have calculated the total LET caused by the nuclear scattering for the 2 MeV α particles of the experiment of Gandikota et al.[20] and for the 1.47 MeV He - 0.84 MeV Li pair of our experiment. In ref.20, the LET is about 27 meV/Å, while in our case the LET is 236 meV/Å, to which Li contributes for about 200 meV/Å. Then, in our method each single event of neutron capture releases to the crystal an energy 9 times larger than a single α particle coming from the accelerator beam of Gandikota et al. [20].

Further, as an index of the effectiveness of damage, we have evaluated the displacement per atom (dpa), which is available in the software for the Stopping and Range of Ions in Matter (SRIM-2003). SRIM-2003 takes as input the results of our Montecarlo simulation code of the two experiments (neutron and α particle irradiation), where the He and Li ions generation points, directions and energies are produced. It results that the ratio of the averaged values of the dpa over the film volume for ours and Gandikota et al.[20] experiments is about 10, in good agreement with the previous LET calculations.

Then, in order to compare the two irradiation methods, we assume as scaling factor the ratio of the average dpa per unit of the fluence of external particles, i.e., neutrons and α particle, considering also the probability of neutron capture by the $^{10}$B, which takes into account the cross section of this reaction. In figure 3, our data on critical temperature and resistivity, along with those from ref.20 are reported as a function of the fluence multiplied by the calculated scaling factor. Despite the difference between the two experiments, a similar behavior of ρ and $T_c$ can be observed. This is a clear indication that the charged particles produced by capture of thermal neutrons and not fast neutrons are mostly responsible for the damage in our films, as already suggested in ref.12 and 14 for bulk samples. In the inset of figure 3, the critical temperatures as a function of resistivity for the two experiments are plotted; both curves show a linear behavior and $T_c$ vanish almost at the same resistivity value (about 100 μΩcm), as already reported in literature for several different series of samples[7,14]. The data of figure 3 clearly indicate that we developed a reliable method to insert disorder in the system: in this set of samples we are able to study physical properties that closely depend on disorder as, for example, normal state magnetoresistivity and critical fields. This will be done in the next paragraphs where a comparison between the scattering times in the two bands coming from the two different analyses will be attempted.

**IV. Normal State Magnetoresistivity**

The normal state magnetoresistivity is a transport property that can be strongly affected by disorder. As discussed in ref. 25 and 26, according to the theory of cyclotron orbits, positive transverse magnetoresistivity is expected in a metal where two or more bands cross the Fermi level. Indeed, the contributions to magnetoresistivity of different bands either cancel or add to each other depending on whether the charge carriers are of the same sign or of opposite sign, respectively. In magnesium diboride, large normal state magnetoresistivity is accounted for by the presence of four bands, namely two π-bands, one of which has electron-like charge carriers and the other hole-like ones, and other two hole-type σ-bands.

Furthermore, the transverse magnetoresistivity of epitaxial films is sensitive to the orientation of the magnetic field with respect to the crystalline axes: depending on the ratio of the scattering times in the two types of bands $\beta=\tau_\pi/\tau_\sigma$, the magnetoresistivity in the configuration with in plane magnetic field can be larger ($\beta>1.6$) or smaller ($\beta<1.6$) than the magnetoresistivity in the configuration with out-of-plane magnetic field. In the approximation of diagonal anisotropic mass tensor, the values of the scattering times $\tau_\pi$ and $\tau_\sigma$ can be extracted by fitting experimental data, thus providing a powerful tool to study the effects of disorder.

Since magnetoresistivity decreases with increasing disorder, high magnetic fields are needed and only the samples with rather low resistivity can be measured.

A quantitative analysis of anisotropic normal state magnetoresistivity has been carried out on IRR0, 10, 15, 20, 25, 30, where an appreciable variation of normal state resistivity with magnetic field was still present applying magnetic field up to 28 T. Figure 4 shows the curves at 42 K with field parallel and perpendicular to the *ab* planes for all these samples up to 28T; some of them (IRR10, 20 and 30) were already reported, up to 45 T, in ref. 25 and have been replotted here for easier comparison in the same range of field. The magnetoresistivity, as expected, is strongly anisotropic and monotonically decreases upon irradiation as residual resistivity and critical temperature. The unirradiated sample has a large magnetoresistivity, about 70% at highest field in the perpendicular direction and 10% in the parallel direction. In the most disordered sample considered here, IRR30, it reduces to 1.5% at the same field in both orientations. A crossover between the curves measured with magnetic field perpendicular and parallel to the *ab* planes occurs passing from IRR10 to IRR25 and finally no difference between the two directions is observed in IRR30. The scattering times in π- and σ-bands obtained from the fit are reported in Table II. These data indicate that the unirradiated film has slightly dirtier π bands, while upon irradiation, both bands become increasingly disordered and eventually have comparable scattering times in the most irradiated samples. This picture seems to be consistent with the nature of irradiation, which should not have different effects on the two types of bands, differently from chemical substitutions. The residual resistivity calculated from these scattering times, shown again in Table II, despite some small differences with respect to the measured values, reproduces quite well the general trend, giving a confirmation of the reliability of this analysis.

The availability of the scattering times in the two bands allows to discuss the evolution with disorder of the conduction regime in the two bands. The regime of conduction can be clean or dirty whether the electron mean free path is longer or shorter than the BCS coherence length ($\ell/\xi_0 \gg 1$ or $\ell/\xi_0 \ll 1$, respectively). In the case of MgB$_2$ we have two mean free paths $\ell_{\sigma,\pi} = v_{F\sigma,\pi}\tau_{\sigma,\pi}$ and two coherence length $\xi_{0\sigma,\pi} = \frac{\hbar v_{F\sigma,\pi}}{\pi \Delta_{\sigma,\pi}(0)}$ where $v_{F\sigma,\pi}$ is the mean in-plane Fermi velocity ($v_{F\sigma}$~4.5×10$^5$ m/s, $v_{F\pi}$~5.6×10$^5$ m/s) and $\Delta_{\pi,\sigma}(0)$ are the energy gaps which can be estimated by the data of Putti et al. from specific heat measurements on irradiated bulk samples at the corresponding critical temperatures[15]. For the most disordered films, where the scattering times from magnetoresistivity were not available, i.e. for IRR35-IRR45 samples, we assume $\tau_\sigma=\tau_\pi=\tau$ and estimate it from resistivity by $\tau^{-1} = \varepsilon_0\rho(\omega_{p\pi}^2 + \omega_{p\sigma}^2) = \varepsilon_0\rho\omega_p^2$, where $\omega_p$ is the plasma frequency defined by $\omega_p^2 = (\omega_{p\sigma}^2 + \omega_{p\pi}^2)$. This assumption is reasonable because already IRR30 has $\tau_\sigma$~$\tau_\pi$ (see Table II).

The $\ell_{\sigma,\pi}/\xi_{0\sigma,\pi}$ ratios of σ and π band for the whole series of samples are summarized in Table III. From these data some conclusions can be drawn. First, the unirradiated film has both bands in clean limit or near, but the π band is more disordered than σ one, being $\ell/\xi_0$ ratio almost 5 times lower. Then, irradiation progressively introduces disorder in both bands and already IRR25, which has a T$_c$ of about 38K, has both bands in dirty regime. Finally, the most irradiated films are uniformly disordered with a very low $\ell/\xi_0$ ratio. These results show that neutron irradiation has a similar effect on both bands, that scales with neutron fluence.

**V. Upper critical field**

In figure 5, upper critical field as a function of temperature for most of the irradiated samples are reported. For IRR15 and IRR25 only the point at 28 T was measured at GHMFL. The *c* axis orientation of the samples allows measurements with magnetic field applied both perpendicular (open symbols, upper panel) and parallel (full symbols, lower panel) to the *ab* planes; for comparison, also data of the unirradiated film, IRR0, have been plotted. The critical fields at zero temperature $H_{c2}(0)$ for all the measured samples obtained by linear extrapolation from the experimental data are reported in the inset of figure 5 as a function of $T_c$.

First, it is worth noticing that the upper critical field of the unirradiated film (7T at 5K in the perpendicular direction) is considerably higher than that of single crystals[27], even though the resistivity is similar. Moreover, a large spread in $H_{c2}$ values of clean films grown by HPCVD is present, as shown by data in the inset of figure 5, where the $H_{c2}(0)$ of three clean samples have been plotted together with the series of irradiated films. Squares are referred to a HPCVD film carefully stored but measured after few weeks since the deposition, whereas rhombs and triangles have been extrapolated from Zeng et al.[18] and from Iavarone et al.[28] respectively. A scatter in $H_{c2}(0)$ in both directions is clearly evident, with values of about 22 and 40T perpendicular and parallel to the planes respectively. This means that aging is able to increase considerably upper critical field, and therefore a certain "reference" $H_{c2}$ is lacking. We have to take this fact into account when we introduce a controlled amount of disorder by irradiation.

Concerning the irradiated samples, an interesting feature is present: in the films where $T_c$ remains nearly unchanged but ρ increases by a factor 10, the upper critical field is almost constant, reaching about 25T at 5 K with B perpendicular to the planes. At the highest irradiation level instead, when $T_c$ decreases, also $H_{c2}$ is hardly suppressed. From the $H_{c2}(0)$ in the inset, this behavior is even more apparent: $H_{c2}(0)$ reaches similar values of about 50T in the parallel direction and about 30 T in the perpendicular one in all the samples with $T_c$ close to the optimal value decreasing when the critical temperature decreases. At 17K an isotropic value was measured. Similar features, even with lower values of $H_{c2}(0)$ (35T in a sample with $T_c$ of 36K) were found in the films damaged by alpha particles of ref. 21, where measurements up to 9T were reported. It is worth noting that, due to the positive curvature of the $H_{c2}$ vs. T curves near $T_c$, disappearing at high fluence, a linear extrapolation from the experimental points at fields lower than 10 T strongly underestimates $H_{c2}(0)$ and leads to a different shape of the $H_{c2}(0)$ vs. $T_c$ curve. Hence, the use of high magnetic fields is required for a correct definition of $H_{c2}(0)$.

In case of dirty thin films, the upper critical field data can be analyzed in the framework of model proposed by Gurevich[29] for two band superconductors, where the diffusivities in the two bands are important parameters. We applied this model to IRR30, where both bands are surely in dirty limit (see Table III) and the small $T_c$ reduction allows us to assume unchanged coupling constants, using the scattering times from Table II to evaluate the diffusivities; the obtained $H_{c2}$ value in the perpendicular direction is about 6 T, almost five times lower than the measured one. In addition, if we consider the less irradiated samples, where both bands are still in clean limit (see Table III), perpendicular $H_{c2}$ is expected to be about 3 T, as in single crystals[27]. This suggests that the upper critical field in thin films could be related in a complex way to the measured resistivity, not only because of the multiband character of $MgB_2$ [29]. Furthermore, from the data of Figure 5, it is clear that upper critical field is the only property considered here that does not systematically scale in the whole range of neutron fluence. The weak dependence on irradiation in the samples with low resistivity suggests that disorder intrinsically present in thin films plays an important role in determining the high values of upper critical field.

In view of these considerations, we tried to search for another mechanism that could account for high $H_{c2}$ values along with a low resistivity. The presence of stresses induced by the substrate during the growth and the possible degradation or contamination of the sample surface are typical features of thin films. In case of HPCVD samples, it has been shown that the silicon carbide substrate is able to induce a strain which enhances the critical temperature[30]; in addition, they are extremely sensitive to air and humidity. Therefore, it is quite reasonable to hypothesize the presence of a layer in the sample with a different resistivity with respect to the rest of the film. In this scenario, it may occur that the measured resistivity and $H_{c2}$ are indeed related to different portions of the sample and thereby they are not easily related to each other. As a first step, let us suppose that the film is made by two blocks: a thick layer with low resistivity $\rho_0$ and a thinner one with a resistivity gradient varying from $\rho_0$ to a maximum value $\rho_{max}$. The measured resistivity is therefore given by the parallel resistance. Moreover, we can suppose that the critical temperature of the thin layer decreases with increasing ρ. Looking at the $T_c$ vs. ρ curve reported in the inset of figure 3, we hypothesize a linear dependence, with $T_c$ vanishing at ρ=100 μΩcm; in this hypothesis, $ρ_{max}$ is 100 μΩcm. It is apparent that the high resistivity part does not appreciably influence the measured ρ. Moreover, despite each layer of different ρ has also a different $T_c$ according to the $T_c$ vs. ρ curve, the experimental $T_c$ is determined by the highest one. The upper critical field of each layer can be calculated in a single band picture following ref. 31:

$$H_{c2}(0) = 0.693 \frac{dH_{c2}}{dT}\bigg|_{T_c} T_c$$

where

$$\frac{dH_{c2}}{dT}\bigg|_{T_c} \propto \frac{T_c(1+\lambda)^2 N(0)}{\omega_p^2}\left(1+0.75\frac{\xi}{\ell}\right) \propto \frac{T_c(1+\lambda)^2 N(0)}{\omega_p^2}\left(1+\frac{1.50\hbar\varepsilon_0}{3.52 K_B}\frac{\omega_p^2 \rho}{T_c(1+\lambda)}\right)$$

Here, $N(0)$ and λ are the density of states and the electron-phonon coupling constant and $\omega_p$ is the plasma frequency as defined in section IV. This formula includes the crossover between clean and dirty limit through the $\xi/\ell$ ratio. If $\omega_p$, $N(0)$ and λ are constant, assuming that their dependence on disorder is weak, $H_{c2}(0)$ depends on the resistivity and critical temperature varying from layer to layer and related to each other by the experimental $T_c$ vs. ρ curve. Now, the upper critical field of the sample is the maximum among those of the parallel layers; this definition makes $H_{c2}$ independent on the arbitrary shape of the distribution of resistivities throughout the sample.

In figure 6, the calculated $H_{c2}$ curves with B perpendicular to the planes are shown and compared to the corresponding curves of homogeneous samples; in all cases, a simple parabolic dependence on the reduced temperature is assumed. In the upper panel, the case of a clean sample with $T_c$=41K and ρ=0.7 μΩ cm is presented, whereas in the lower panel there is the case of a dirty sample with $T_c$=22K and ρ=36 μΩ cm. In the insets, the assumed spatial distribution of ρ (left hand axis) and $T_c$ (right hand axis) along the sample thickness is shown. The curves of both panels are normalised to the $H_{c2}(0)$ value of the homogeneous sample. It can be seen that in the clean sample $H_{c2}$ is considerably increased by the presence of the resistive layer and an upward curvature near $T_c$ of the $H_{c2}$ vs T curve appears. This means that until the resistivity of the thick layer $\rho_0$ is low enough, the upper critical field is determined by the resistive layer. Only when we start to introduce disorder and $\rho_0$ increases up to a certain value, then $H_{c2}$ is mainly determined by the thick layer, as in the case of the dirtier sample with reduced $T_c$ shown in the lower panel of figure 6, where the contribution of the thin resistive layer is less important, and $H_{c2}$ scales reasonably with irradiation. This simple model is able to explain quite well the typical features observed in the upper critical field of thin films. In fact, it can account for the high values of critical fields in low resistivity samples, as well as for their weak dependence on resistivity at low irradiation levels and it reproduces well the upward curvature near $T_c$. Moreover, it reproduces also the trend of the more irradiated films, where $H_{c2}$ depends on disorder, which suppresses the critical temperature and increases $\rho$. Obviously, to have an accurate fit of the critical field curves the two band nature of $MgB_2$ has to be taken into account: this is out of the purposes of this article and will be done extensively in a future work.

## VI. Conclusions

We have carried out a systematic study of the effect of disorder in neutron irradiated $MgB_2$ thin films. We reached a good control of the damage process, obtaining a set of samples where several properties, such as critical temperature, residual resistivity and normal state magnetoresistivity scale well with increasing neutron fluence. We evidenced how it is possible to follow the passage from clean to dirty limit in each band by the analysis of magnetoresistivity measurements. Upper critical field is the only property that does not seem to depend on irradiation, at least at low fluences and therefore the independence of the upper critical field on resistivity is confirmed. We tried to find a possible explanation other than the two band nature of $MgB_2$ proposing a simple model based on the presence of a layer with high resistivity in parallel to the film. When the sample is clean, the measured $H_{c2}$ is determined by this layer, that only slightly affects the measured resistivity and magnetoresistivity. When the resistivity of the film increases up to a certain limit, i.e. upon irradiation, then the resistive layer becomes less important and $H_{c2}$ scales well with ρ. Despite the model has to be implemented to take into account the multiband nature of magnesium diboride, we showed that it can account for an increase of the upper critical field especially in samples with low resistivity, just the typical phenomenology observed in thin films, as well as for the trend upon irradiation.

**Acknowledgements**

We acknowledge R.Vaglio for the useful discussions on upper critical field and the support of the European Commission from the 6th framework programme "Transnational Access - Specific Support Action", contract N° RITA-CT-2003-505474. This work is partially supported by the project PRIN2004. The work at Penn State is supported in part by NSF under Grant No. DMR-0306746 and by ONR under Grant No. N00014-00-1-0294

**Tables**

**Table I**: List of the main properties of the complete series of irradiated samples: neutron fluence, critical temperature (defined as 50% of residual resistivity) and transition widths in K, residual resistivity ratios, residual resistivities $\rho_{exp}$ (calculated at 42 K) and $\Delta\rho=\rho(300K)-\rho_0$ in μΩcm, resistivity normalized using $\Delta\rho$=7.6 μΩcm , $\rho_{corr}$, in μΩcm and *c* axes in Angstroms.

| Sample | Neutron Fluence ($cm^{-2}$) | $T_c$ (K) | $\Delta T_c$ (K) | RRR | $\rho_{exp}$ (μΩ·cm) | $\rho_{corr}$ (μΩ·cm) | $\Delta\rho$ (μΩ·cm) | c axis (Å) |
|---|---|---|---|---|---|---|---|---|
| **IRR 0** | -- | 41.1 | 0.1 | 16 | 0.55 | 0.7 | 9 | 3.508 |
| **IRR 10** | $6.4 \cdot 10^{15}$ | 41.05 | 0.1 | 9.8 | 1.0 | 0.9 | 8.8 | 3.513 |
| **IRR 15** | $4.1 \cdot 10^{16}$ | 40.3 | 0.4 | 7 | 2.7 | 1.7 | 12 | |
| **IRR 20** | $6.5 \cdot 10^{16}$ | 40.7 | 0.2 | 4.5 | 2.3 | 2.2 | 8 | 3.515 |
| **IRR 25** | $4.1 \cdot 10^{17}$ | 38.3 | 0.4 | 2.4 | 8.6 | 6 | 11 | |
| **IRR 30** | $7.6 \cdot 10^{17}$ | 36.1 | 0.4 | 1.6 | 16 | 12 | 10 | 3.531 |
| **IRR 35** | $3.0 \cdot 10^{18}$ | 22.2 | 1.9 | 1.2 | 90 | 36 | 19 | 3.553 |
| **IRR 36** | $3.2 \cdot 10^{18}$ | 30.1 | 2.3 | 1.3 | 58 | 27 | 16 | 3.540 |
| **IRR 40** | $9.5 \cdot 10^{18}$ | 17 | 4.0 | 1.1 | 69 | 55 | 9.5 | 3.549 |
| **IRR 41** | $1.0 \cdot 10^{19}$ | 13.3 | 0.6 | 1.1 | 82 | 57 | 11 | 3.548 |
| **IRR 45** | $2.5 \cdot 10^{19}$ | 2.75 | 0.2 | 1 | 560 | 125 | 34 | 3.551 |
| **IRR 50** | $1.1 \cdot 10^{20}$ | - | - | - | 2585 | 197 | 100 | 3.543 |

**Table II**: Scattering times of σ and π bands estimated from magnetorsistivity ($\tau_{\sigma magres}$ and $\tau_{\pi magres}$), as well as resistivities calculated by these values of τ ($\rho_{magres}$) for IRR0-30. In the last column, the corrected resistivity $\rho_{corr}$ has been also reported for comparison.

| Sample | $\tau_{\pi\,magres}$ (s) | $\tau_{\sigma\,magres}$ (s) | β | $\rho_{magres}$ (μΩ·cm) | $\rho_{corr}$ (μΩ·cm) |
|---|---|---|---|---|---|
| **IRR 0** | $8.9 \cdot 10^{-14}$ | $1.8 \cdot 10^{-13}$ | 0.5 | 0.6 | 0.7 |
| **IRR 10** | $9.9 \cdot 10^{-14}$ | $1.6 \cdot 10^{-13}$ | 0.6 | 1.2 | 0.9 |
| **IRR 15** | $8.9 \cdot 10^{-14}$ | $6.4 \cdot 10^{-14}$ | 1.4 | 1.9 | 1.7 |
| **IRR 20** | $7.4 \cdot 10^{-14}$ | $3.0 \cdot 10^{-14}$ | 2.5 | 2.7 | 2.2 |
| **IRR 25** | $3.6 \cdot 10^{-14}$ | $1.8 \cdot 10^{-14}$ | 2.0 | 5.1 | 6.0 |
| **IRR 30** | $2.0 \cdot 10^{-14}$ | $1.2 \cdot 10^{-14}$ | 1.6 | 9.2 | 12 |

**Table III**: $\ell/\xi$ ratio for σ and π bands of the whole series of irradiated films.

| Sample | $\ell/\xi$ π band | $\ell/\xi$ σ band |
|---|---|---|
| **IRR 0** | 0.9 | 5.1 |
| **IRR 10** | 1 | 3.6 |
| **IRR 15** | 0.9 | 2 |
| **IRR 20** | 0.8 | 0.9 |
| **IRR 25** | 0.3 | 0.4 |
| **IRR 30** | 0.1 | 0.1 |
| **IRR 35** | 0.03 | 0.03 |
| **IRR 36** | 0.04 | 0.04 |
| **IRR 40** | 0.02 | 0.02 |
| **IRR 41** | 0.02 | 0.02 |
| **IRR 45** | 0.01 | 0.008 |
| **IRR 50** | 0.006 | 0.005 |

**Figure Caption**

**Figure 1**: X-ray patterns around (001) and (002) reflections of $MgB_2$; for simplicity, only selected films are reported.

**Figure 2**: *c* axis parameter, critical temperature and residual resistivity as a function of thermal neutron fluence for the whole series of irradiated films.

**Figure 3**: Resistivity and critical temperature at 50% of the transition as a function of fluence of equivalent particles (main panel) and $T_c$ as a function of resistivity (inset) for our films and for the films from Gandikota et al. (ref. 20).

**Figure 4**: Magnetoresistivity as a function of squared magnetic field in parallel (full symbols) and perpendicular (open symbols) orientation for IRR0-IRR30 samples. The curves of IRR10, 20 and 30 are those reported in ref.25.

**Figure 5**: Upper critical field as a function of temperature measured with magnetic field perpendicular (upper panel, empty symbols) and parallel (lower panel, full symbols) to the *ab* planes. In the inset: $H_{c2}$ extrapolation at 0 K as a function of critical temperature for the samples of this work (circles), for clean films grown by HPCVD from literature (◆ from ref. 18, ▼ from ref. 28) and for a clean film measured after some time since the deposition (■).

**Figure 6**: Calculated $H_{c2}(T)$ curves in case of a resistivity gradient in the film and of a homogeneous film. The cases of a resistivity of 0.7 μΩcm and 36 μΩcm are reported in upper and lower panel respectively. In the insets the ρ and $T_c$ behavior along the film thickness for the two cases.

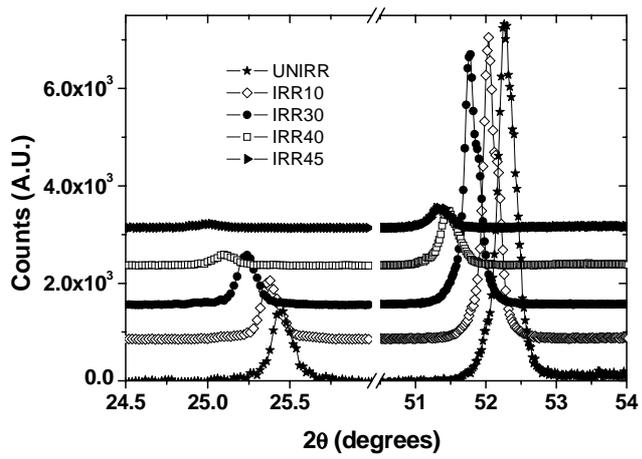

Figure 1

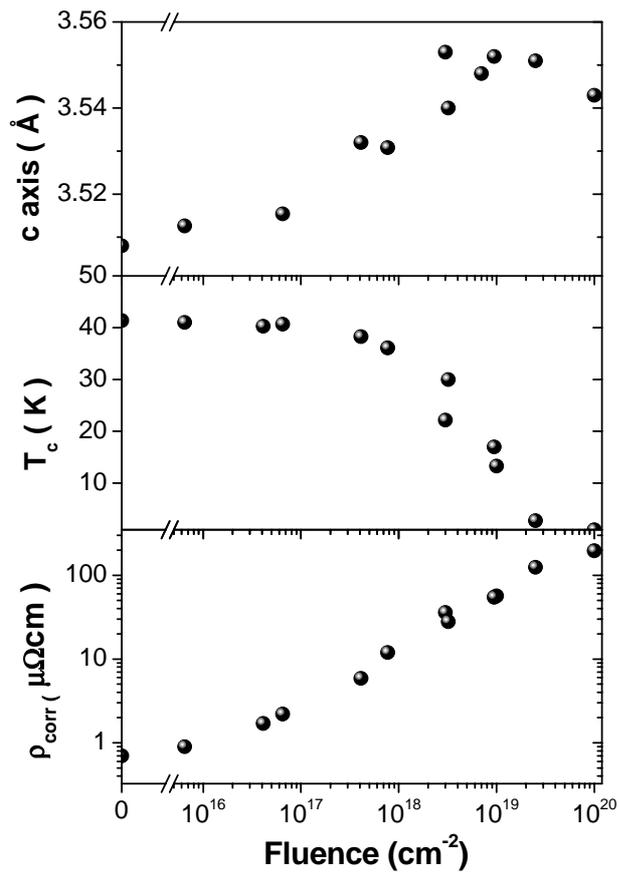

Figure 2

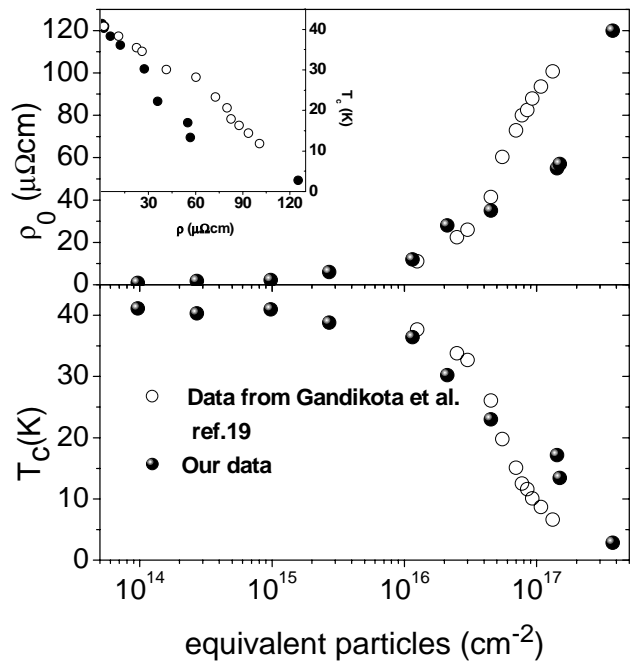

Figure 3

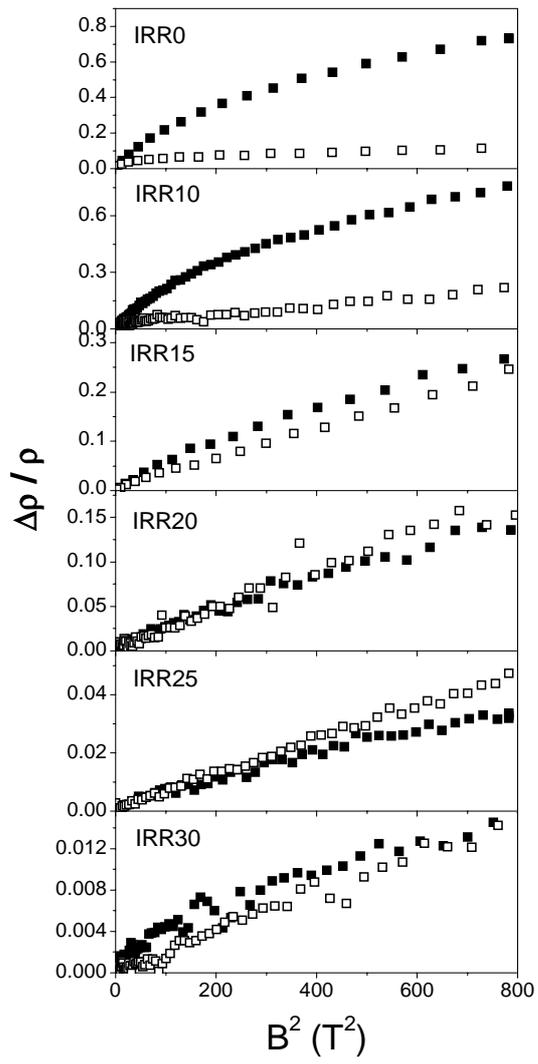

Figure 4

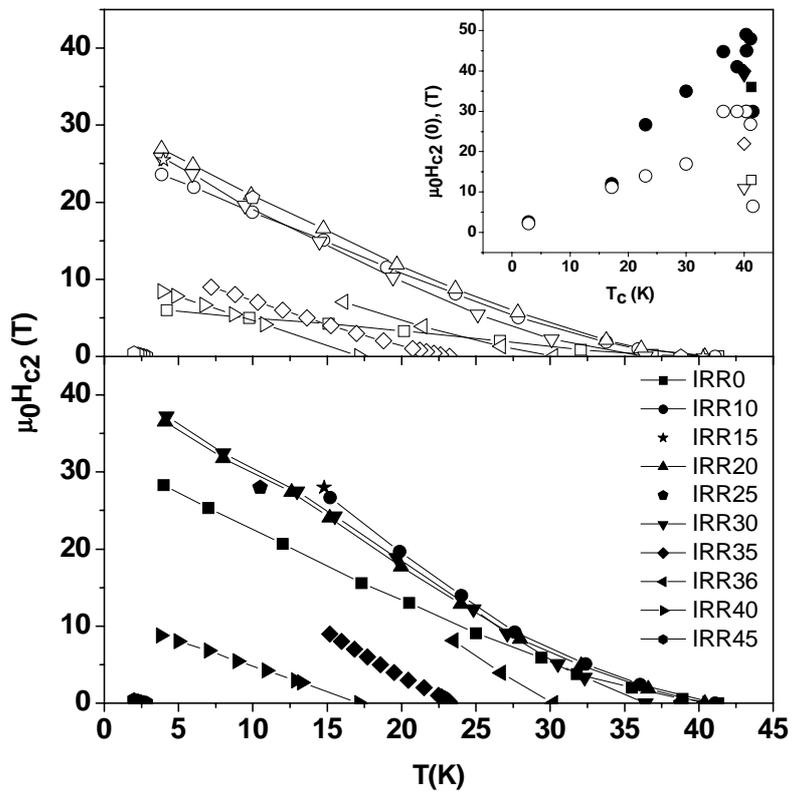

Figure 5

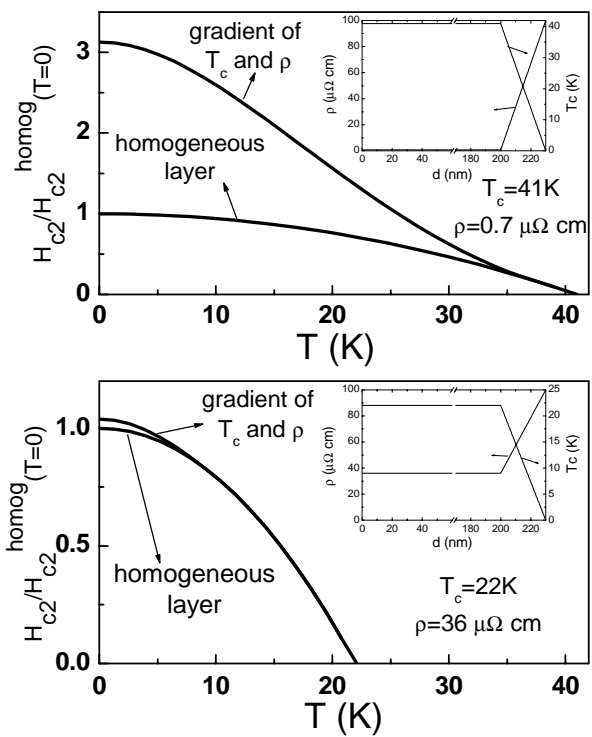

Figure 6